\documentstyle[aps,graphicx,twocolumn,epsf,floats]{revtex}

\begin{document}


\title{Sampling rare events: statistics of local sequence alignments}
\author{Alexander K. Hartmann}
\address{Department of Physics, University of California,
Santa Cruz CA 95064, USA\\
      E-mail: \texttt{hartmann@bach.ucsc.edu}
        }
\date{\today}
\maketitle 

\begin{abstract}
{\em Summary} A new method to simulate probability distributions in regions
where the events are {\em very unlikely} (e.g. $p\sim 10^{-40}$)
is presented. The basic idea is to represent
the underlying probability space by the phase space of a physical system.
The system is held
at a temperature $T$, which is chosen such that the system preferably
generates configurations which originally have low probabilities. 
Since the distribution of such a physical
system is know from statistical physics, the original unbiased distribution
can be obtained.

As an application, local alignment of 
protein sequences based on BLOSUM62 substitution scores with $(12,1)$
affine gap costs are considered.
The distribution of optimum sequence-alignment scores $S$ is studied 
numerically over a large range of scores. 
 The deviation of $p(S)$ from 
the extreme-value (or Gumbel) distribution is quantified. This
deviation decreases with growing sequence length.

\end{abstract}

\paragraph*{Introduction}
Modern molecular biology, e.g.  the {\em human genome project}
\cite{humane-genome}, relies heavily on the use of large 
databases\cite{brown2000,rashidi2000}, where DNA or
protein sequences are stored. The basic tool for accessing these 
databases and comparing different sequences is {\em sequence alignment}.
The result of each comparison is an maximum alignment {\em score} $S$.
One is interested either in {\em global} or {\em local} optimum alignments. For the
first case, the score is maximized over all alignments which include
both sequences fully, while for the second case, 
the optimum is calculated over all
possible subsequences. 
To estimate the significance of the result of a comparison, one has
to know, based on a random model, 
the statistical distribution $p(S)$ of scores.
For biologically relevant models, e.g. for protein sequences with
BLOSUM62 substitution scores\cite{blosum62} and affine gap 
costs\cite{gotoh1982},
$p(S)$ is not known in the interesting region, where $p(S)$ is small.
A number of empirical 
studies\cite{smith1985,collins1988,mott1992,waterman,altschul1996}
 for local alignment, in the region where $p(S)$ is large, suggest
that $p(S)$ is an {\em extreme value} (or Gumbel)  
distribution\cite{gumbel1958}
\begin{equation}
p_G(S)=\lambda e^{-\lambda(S-u)}\exp(-e^{-\lambda(S-u)})\,,
\end{equation} where $u$
denotes the maximum of the distribution and $\lambda$ characterizes the
behavior for large values of $S$, i.e. the tail of the distribution.

In this work, to determine the tail of $p(S)$, 
a {\em rare event simulation}
is applied. For dynamical problems, like investigating 
queuing systems or studying the reliability of technical components,
several techniques\cite{ripley1987}, mostly based on {\em importance sampling} 
have been developed.
Related methods have been introduced in physics, 
like {\em multi-canonical sampling}\cite{berg1992}
 or the {\em pruned-enriched Rosenbluth method}\cite{grassberger1997}.

By simply changing perspective, one can apply these
standard techniques to many other problems.
Here, the method it is applied to the sequence alignment problem.
 Instead of  directly drawing the random
sequences, the basic idea is to use a physical system, which has a state
given by a pair of sequences and is held at temperature $T$. 
This is similar to the simulated annealing approach\cite{kirkpatrick1983},
used to solve hard optimizations problems.
The
state of the system changes in time, governed by the rules of 
statistical mechanics. 
The energy $E$ of the system is defined as $E=-S$. Therefore, at low
temperatures the system prefers a state with
high score value $S$. Since the thermodynamic properties of the system
are known, it is possible to extract from the measured distribution 
$p^{*}(S)$ of scores the target distribution $p(S)$.

For the alignment problem two sequences ${\bf x}=x_1x_2\ldots x_n$
and ${\bf y}=y_1y_2\ldots y_m$ over a finite alphabet with $r$ letters
are given. For DNA the alphabet has 4 letters, representing the 
bases, for protein sequences it has 20 letters, representing the
amino acids. Let $f_i$ be the probability for the occurrence of letter $i$,
assuming here that all letters of a sequence are independent.
An alignment is a pairing $\{ (x_{i_k},y_{j_k})\}$ 
($k=1,2,\ldots,K$, $1\le i_k<i_{k+1}\le n$ and $1\le j_k<j_{k+1}\le m$) 
of letters from the two sequences. Please note that
some letters may not be aligned, i.e. {\em gaps} occur. 
To each alignment a score is assigned,
 via a scoring function $s(x,y)$.
The total score is the sum of scores of all aligned letters 
$\sum_k s(x_{i_k},y_{j_k})$ plus the costs of all gaps.
Here, so called {\em affine} gap costs $(\alpha,\beta)$ 
are considered, i.e. a gap of length $l$ has costs $g(l)=-\alpha-\beta (l-1)$.
For both global and local alignment efficient 
algorithms\cite{needleman1970,smith1981,gotoh1982} exist,
which calculate an optimum alignment in time $O(nm)$. Hence, one can easily
generate e.g. $N\approx 10^{5}$ samples of pair of sequences according
the frequencies $f_i$, obtain
each time the optimum alignment, and calculate a histogram of the
scores $S$. This {\em simple sampling} allows one to
calculate $p(S)$ in the region were the probabilities are large (e.g.
$p(S)\approx 10^{-4}$). Recently, the {\em island method}
\cite{altschul2001} was
introduced, which allows a speed up of several orders of magnitudes,
but still the far end of the distribution is out of reach.

\paragraph*{Algorithm}

The basic idea to determine the behavior of $p(S)$
 at the rare event tail (e.g. $p(S)\approx 10^{-40}$) is 
to view a pair of sequences as a physical system, which
behaves according the rules of statistical mechanics.
This means that instead of 
considering many independent pairs of fixed sequences,
one pair of sequences $c(t)=({\bf x}(t), {\bf y}(t))$  is used,
which changes at discrete times $t$.
Mathematically speaking, a Markov chain\cite{lipschutz2000}
 $c(0)\rightarrow c(1) \rightarrow c(2) \rightarrow \ldots$ is used
to generate the instances.
The behavior of a Markov chain is not deterministic, but governed by 
probabilities $p(c\rightarrow c^{'})$ that state $c^{'}$ follows immediately
after state $c$.

The simplest rule for the transition is, to choose randomly a position
in one of the sequences and choose randomly a new letter from the alphabet,
with all positions being equiprobable and the letters having
probabilities $f_i$, i.e. 
$p(c\rightarrow c^{'})=f_i/(n+m)$ if $c,c^{'}$ differ by at 
most one letter, and $p(c\rightarrow c^{'})=0$ otherwise. With this choice
of the transition probabilities, for $t\to \infty$ all possible pairs
of sequence have the  probability $P(c)=\prod_i f_{x_i}\prod_j f_{y_j}$ 
of occurring. Hence, simple sampling is reproduced.

To increase the efficiency, one can apply {\em importance sampling}
\cite{ripley1987,metropolis1953}, a standard method for simulating rare
events\cite{kalashnikov1997}, which allows one to concentrate the sampling in
small regions in configuration space. Suppose, we want to calculate the
expectation value $\langle g \rangle_P$ of a function $g$ on a (here discrete)
configuration space $\{c\}$, with respect to the probabilities $P(c)$.
Thus, we have $\langle g \rangle = \sum_{c} g(c) P(c)$. To estimate
$\langle g \rangle_P$, we
generate $N$ samples $c^{(i)}$ which are distributed according to $P$, 
and calculate $1/N \sum_i g(c^{(i)})$.
Let now $Q(c)$ be another
probability distribution with $Q(c)\neq 0 $ for all $c$. Then we can write
$\langle g \rangle_P = \sum_{c} \frac{g(c)P(c)}{Q(c)}Q(c) = 
\langle gP/Q \rangle_Q$. This means,
when we generate N configurations $c^{(i)}$ according to 
the probabilities $Q(c)$,
then $1/N \sum_i f(c^{(i)})P(c^{(i)})/Q(c^{(i)})$ is also an estimator for 
$\langle f \rangle_P$. By choosing $Q$ such that the sampling occurs in
the region of the phase space we are interested in, we can obtain a
much higher accuracy there.

A good choice for the transition rule of the sequence-alignment problem is 
to select one position in the pair $c$ of sequences randomly,
change the letter randomly according the frequencies $f_i$,
recalculate the optimum alignment $S(c^{'})$ with a standard algorithm
and accept this move $c\to c^{'}$ with probability 
$\max(1,\exp(\Delta S/T))$, where $\Delta S=S(c^{'})-S(c)$.
This transition probabilities describe a  physical system 
at temperature $T$ with energy $E=-S$. The advantage of this approach is
that the equilibrium distribution $Q(c)$  is known
from statistical physics\cite{reichl1998}: $Q(c)=P(c)\exp(S(c)/T)/Z$ with
$Z(T)=\sum_c P(c)\exp(S(c)/T)$ being a normalization constant, called the
partition function. Thus, the probability to have score $S$ is
\begin{equation}
p^{*}(S)=\frac{\exp(S/T)}{Z(T)}{\sum_c}^{'} P(c)\,,
\end{equation} 
where the sum ${\sum}^{'}$ runs over all sequences with score $S$.
Thus, to obtain $p(S)$, one simulates the system at temperature $T$
and records a histogram of scores $p^*(S)$. Then, the unbiased estimator
for the distribution is $p(S)={\sum_c}^{'} P(c)=p^*(S)Z(T)\exp(-S/T)$. 
Note that $Z(T)$ is unknown a priori, but can be determined
very easily, as shown below.

To describe the behavior of $p(S)$ over a wide range, the
model must be simulated at several temperatures. For this reason, and
because at low temperatures the dynamics are slow in general so we
want to increase the efficiency, the model
is simulated via the {\em parallel tempering} 
method\cite{marinari1992,hukushima1996}. 
Using this technique, the system is simulated at $N_T$ 
different temperatures $T_0<T_1<\ldots<T_{N_T}$
in parallel, i.e. with $N_T$ independent pairs $c(T_i)$ of sequences. 
The main idea of parallel tempering is, that from time to time 
the configurations between neighboring
temperatures $T_i$, $T_{i+1}$ are exchanged according a probabilistic
rule\cite{marinari1992,hukushima1996}. Here,
each simulation step consists of one Markov  step for each configuration
$c$ and one exchange step between one neighboring pair $c(T_i),c(T_{i+1})$.

\paragraph*{Example}

\begin{figure}[htb]
\begin{center}
\epsfxsize=\columnwidth
\epsfbox{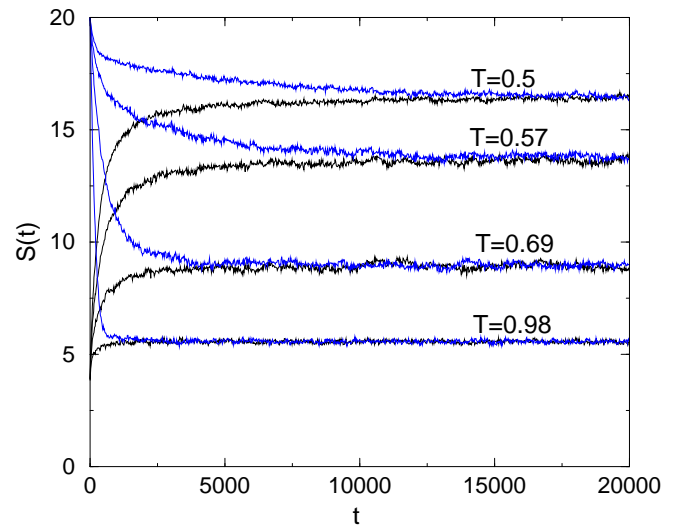}
\caption{Average alignment score $S$ as a function of step $t$ 
for $n,m=20$, 4 letters, 
local alignment without gaps for different temperatures $T$. 
For each temperature, 1000 simulations were started with two random sequences
(low scores) and 1000 simulations with two equal sequences (high scores).}
\label{figEt}
\end{center}
\end{figure}

Next, a simple example is given, illustrating how the method works. 
Optimum local alignments without gaps for sequences of equal length  $m=n=20$ 
and $r=4$ letters, all having the
same probability $1/4$, are calculated. For the test the following score 
is applied:  $s(x,y)=1$ if $x=y$ and
$s(x,y)=-3$ otherwise. Two types of runs are performed:
(a) initially, all pairs of sequences are random, and (b) each pair consists
of two equal sequences. Thus, for the first type, initially the score
is low, while for the second type the score is initially maximal. 
This provides a 
criterion for equilibration: if the average score for both types
of initial configuration agree within error bars (at time $t_0$), 
the simulation is long enough. In Fig.
\ref{figEt} the average optimum score $S$ from 1000 independent runs 
and four different temperatures $T$ is shown.

To obtain uncorrelated samples, only samples at $t_0$, $t_0+\tau$, $t_0+2\tau$
etc are taken, where $\tau$ is the characteristic time in which
the score-score correlation 
\begin{equation}
c_S(t_0,t) = \frac{\langle S(t_0)S(t)\rangle -\langle S\rangle^2} 
{\langle S^2 \rangle - \langle S \rangle ^2}
\end{equation}
decreases to $1/e$.

\begin{figure}[htb]
\begin{center}
\epsfxsize=\columnwidth
\epsfbox{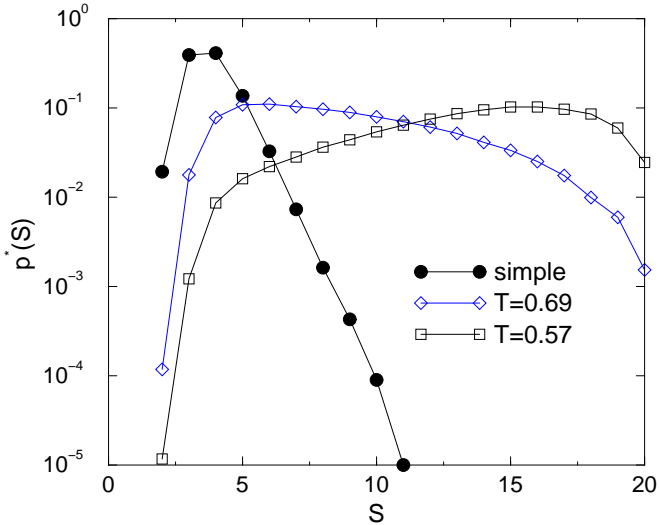}
\caption{Raw distribution of alignment scores $S$ for the direct simulation
and for $T=0.57$ and $T=0.69$
with $n=m=20$, 4 letters and local alignment without gaps.}
\label{figPS}
\end{center}
\end{figure}

In Fig. \ref{figPS} the raw distributions of $S$ for two temperatures is shown
together with a distribution from a simple sampling of $N=10^4$
realizations. Clearly, with the
statistical mechanics approach, the region of high scores is sampled much
more frequently.

\begin{figure}[htb]
\begin{center}
\epsfxsize=\columnwidth
\epsfbox{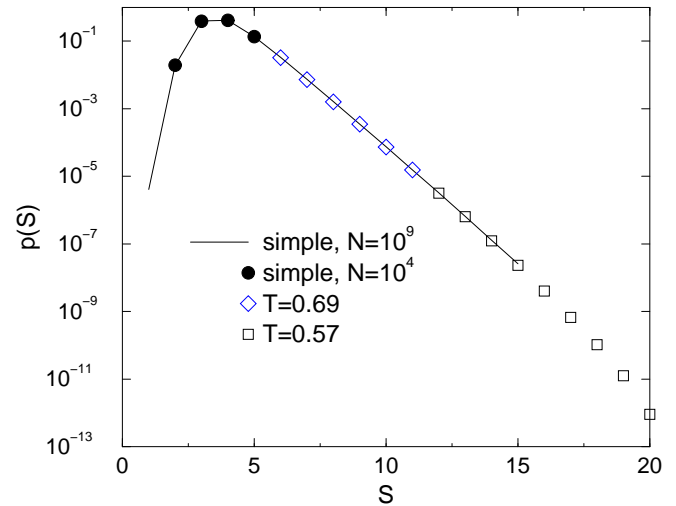}
\caption{Rescaled distribution  $p(S)$ 
for the direct simulation and for $T=0.57$, $T=0.69$
for $n,m=20$, 4 letters, local alignment without gaps. The solid line is the
result of a large simple sampling simulation with $N=10^9$ samples.}
\label{figPSfinal}
\end{center}
\end{figure}

For low scores, the final 
distributions obtained from the simple sampling and from
the finite-temperature simulation must agree. This can be used to
determine the constant $Z(T)$. It is chosen such that the
difference in an interval $[S_1,S_2]$ between the simple sampling 
distribution and the rescaled distribution at $T$ is minimal. In the same
way the constants at lower temperatures can be obtained by matching
to distributions obtained before at higher temperatures. The
final distribution is shown in Fig. \ref{figPSfinal}. For each datapoint,
the distribution with the highest accuracy was taken.
For comparison, a
simple sampling distribution obtained using a huge number of samples 
($N=10^9$) is shown. Both
results agree very well. Please note that the distribution from the
finite-$T$ approach spans almost the entire interval $[0,20]$. 
In principal, the region for very small score $S$
can be investigated also using the new method by simulating at {\em negative}
temperature. How powerful the new method is can be seen
by looking at the right
border of the interval, where a value $p(20)=9.13(20)\times 10^{-13}$ was
obtained. This agrees within error bars with the exact result\cite{exact} 
$0.25^{20}\approx 9.09\times 10^{-13}$. 
This example illustrates, that the method
presented here is indeed able to calculate accurately 
the distribution $p(S)$ of optimum alignment scores in regions where
$p(S)$ is very small.

\paragraph*{Results}

Next, the results for a biologically relevant case are presented.
Sequences of amino acids distributed according the background
frequencies by Robinson and Robinson\cite{robinson1991} are used 
together with the BLOSUM62 scoring matrix\cite{blosum62} for $(12,1)$
affine gap costs. This type of system has been studied in Ref. 
\onlinecite{altschul1996} in the region where $p(S)$ is large. Here,
sequences of length $n=m$ in the range $[40,400]$ were considered. The
simulations were performed for $n_T=7$ temperatures $T\in [2\ldots 10]$
($[3.5\ldots 10]$ for $n,m=400$), with
up to 100 independent runs of length up to $t_{\max}= 4\times10^5$ steps.
For the lowest temperatures it was not possible to equilibrate the system
within the given time. The reason is that near $T\approx 1/\lambda$ the
equilibration times seem to diverge. This indicates a phase transition 
in the physical system with (probably) a glassy phase at low temperatures.
Hence, for the evaluation the only temperatures
used, were those where equilibration was possible.

\begin{figure}[htb]
\begin{center}
\epsfxsize=\columnwidth
\epsfbox{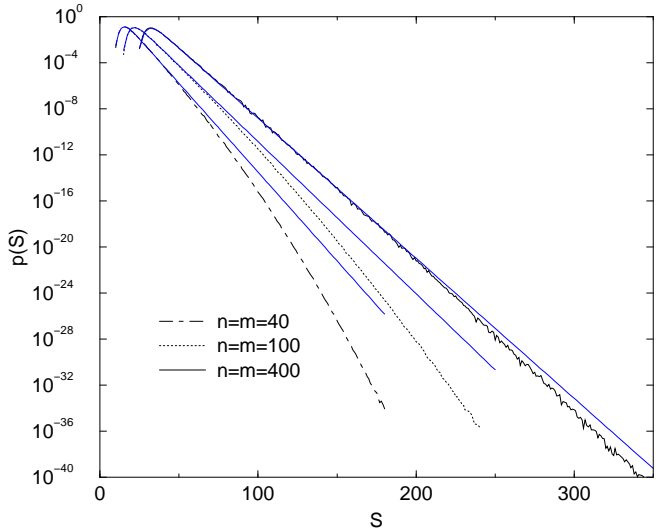}
\caption{Distribution of alignment scores $S$ 
for $L=40,100,200,400$, BLOSUM62 substitution matrix and affine (12,1) gap
costs. The thin solid lines are fits to 
extreme-value distributions with parameters
($\lambda,u$), yielding $(\lambda,u)=(0.355(5),15.35(4))$ ($n,m=40$),
$(0.304(2),21.67(4))$ ($n,m=100$) 
and $(0.280(3),32.01(3))$ for $n,m=400$.}
\label{figPSblosum}
\end{center}
\end{figure}

In Fig. \ref{figPSblosum} the distributions $p(S)$
of optimum alignment scores obtained in this way 
are shown. It was possible to determine the
distribution in regions where $p(S)$ is as small as $10^{-40}$. To obtain the
same accuracy with a simple-sampling approach, given a computer which
optimizes say $10^6$ samples per second, a total simulation time
of about $2.5\times 10^{17}$ times the age of the universe would be necessary.
Also shown in Fig. \ref{figPSblosum}
 are fits of the low-score
data to Gumbel distributions. The resulting parameters ($\lambda,u$)
are comparable to the values found\cite{altschul1996} before and
depend slightly on the sequence length. 
For high scores, significant deviations are visible, in contrast to the 
earlier predictions. Since the deviations occur at high score values,
they could not be detected before using conventional methods.
The reason for the deviations is edge effects: 
very long alignments cannot start near the end of either
of the sequences, so they become even more unlikely. 
The results found here can be fitted very well to 
{\em modified} Gumbel distributions of the form
\begin{equation}
\tilde{p}_G(S)=k\lambda e^{-\lambda(S-u)-\lambda_2(S-u)^2}
\exp(-e^{-\lambda(S-u)})\,,
\end{equation}
with $k\approx 1$, resulting in values for $(\lambda,\lambda_2,u)$ of
$(0.3277\pm0.0003,8.56\times10^{-4}\pm 3\times10^{-6},15.35\pm0.04)$ 
for $n,m=40$,
$(0.2783\pm0.0003,1.72\times10^{-4}\pm 1\times10^{-6},21.67\pm 0.04)$ 
for $n,m=100$, and
$(0.2733\pm0.0004,6.1\times 10^{-5}\pm 2\times 10^{-6},32.01\pm 0.03)$ 
for $n,m=400$.
Anyway, with increasing lengths $n,m$, on a scale of scores proportional to 
$u\sim \log n$, $p(S)$ approaches the Gumbel distribution more and more,
i.e. $\lim_{n\to\infty}\lambda_2=0$.

\paragraph*{Summary}
A method has been presented which allows one to study rare events in random
systems down to regions of {\em very low} probabilities.  
The basic idea is to represent
the probability space by the phase space of a physical system held at
temperature $T$. From the  distribution of states,
the original unbiased distribution can be obtained.

Here, the method is applied to the local sequence-alignment problem.
A biologically relevant case is treated. The distribution of optimum
alignment scores can be studied in regions where the probability is
as small as $10^{-40}$, and yet the deviations of the distribution from the
theoretical prediction are visible. However, with increasing sequence lengths,
the distribution indeed approaches  the Gumbel distribution. Hence, for very
long sequences, when one is not interested in the deviation from the
Gumbel form, the island method\cite{altschul2001} is a more suitable
tool to study sequence alignment. Please note that the island method
is not a general purpose method like the technique presented here, but
designed especially for local alignments.

\paragraph*{Acknowledgements}

The author developed the idea for this method at the workshop 
``Statistical Physics
of Biological Information'' at the  Institute for Theoretical Physics
in Santa Barbara during discussions with P. Grassberger and E. Marinari. 
The author would like to thank A.P. Young and P. Grassberger 
for critically reading the manuscript and interesting discussions
and  A.P. Young also for various other support.
The simulations were performed on a Beowulf Cluster
at the Institut f\"ur Theoretische Physik of the Universit\"at
Magdeburg (Germany). 
Financial support was obtained from the DFG (Deutsche 
Forschungsgemeinschaft)
under grant Ha 3169/1-1.

\end{document}